\documentclass[aps,prl,twocolumn,showpacs,floatfix]{revtex4}

\usepackage{amsmath}
\usepackage{amssymb}
\usepackage{graphicx}
\usepackage{epstopdf}
\usepackage{verbatim}
\usepackage{color}
\usepackage{appendix}
\usepackage[colorlinks, linkcolor=blue, citecolor=blue, urlcolor=blue,breaklinks=true]{hyperref}
\usepackage{soul}
\usepackage{dsfont}

\newcommand{\be}{\begin{equation}}
\newcommand{\ee}{\end{equation}}
\newcommand{\bsubeq}{\begin{subequations}}
\newcommand{\esubeq}{\end{subequations}}

\def\ud{\mathrm{d}}
\newcommand{\bmx}{\begin{pmatrix}}
\newcommand{\emx}{\end{pmatrix}}
\newcommand{\bsmx}{\begin{smallmatrix}}
\newcommand{\esmx}{\end{smallmatrix}}
\newcommand{\vect}[2]{\begin{pmatrix} {#1} \\ {#2} \end{pmatrix}}

\begin{document}

\title{The Open-System Dicke-Model Quantum Phase Transition with a Sub-Ohmic Bath}
\author{D. Nagy}
\author{P. Domokos}

\affiliation{Institute for Solid State Physics and Optics, Wigner Research Centre, Hungarian Academy of Sciences, H-1525 Budapest P.O. Box 49, Hungary}
\begin{abstract}
We show that the critical exponent of a quantum phase transition in a
damped-driven open system is determined by the spectral density
function of the reservoir. We consider the open-system variant of the
Dicke model, where the driven boson mode and also the large N-spin
couple to independent reservoirs at zero temperature. The critical
exponent, which is $1$ if there is no spin-bath coupling, decreases
below 1 when the spin couples to a sub-Ohmic reservoir.
\end{abstract}

\pacs{05.30.Rt,42.50.Pq,37.10.Vz,37.30.+i} 

\maketitle

Quantum critical behavior  appears in driven dissipative systems when the steady-state \cite{Morrison2008Dynamical,Nagy2008Selforganization,Piazza2014Quantum}, rather than the ground state of a Hamiltonian \cite{sachdev2011quantum,Strack2011Dicke,Piazza2014Umklapp}, undergoes a non-analytic, symmetry-breaking change at a critical parameter value. The interplay of an external coherent excitation and  the dissipation can lead to a steady-state which is far from the ground or thermal state. Driven dissipative systems cannot be, in general, mapped onto an effective Hamiltonian system.  It is thus unclear how the critical behavior of the open system is related to universality classes of known quantum and thermal phase transitions  \cite{DallaTorre2010Quantum}.

Recent observation of the Dicke-model superradiant phase transition motivates us to raise this question. Ultracold atoms coupled to the radiation field of an optical resonator allowed for the quantum simulation of the Dicke model  \cite{Dimer2007Proposed,Nagy2010DickeModel,baumann2010dicke} and the experimental demonstration of the phase transition \cite{baumann2010dicke,Baumann2011Exploring,mottl2012roton,Baden2013Realization,Kessler2014Steering,Schmidt2014Dynamical,Klinder2015Dynamical}. The boson component of the model is represented by a single mode of a high finesse optical cavity. The spin-$N$ component is effectively realized by constraining the motion of ultracold atoms into the space of two momentum eigenstates. The interaction is implemented by  a far-detuned laser field illuminating the atoms from a direction perpendicular to the cavity axis. Photons scatter into the cavity, which has a recoil on the atoms. Above a threshold of the laser intensity, which translates to the coupling strength between the spin and the boson mode, a mean field of the cavity mode and the large spin is formed spontaneously.  

Since the cavity mode is coupled through the mirrors to the outside
electromagnetic vacuum field, this system is intrinsically open. It is
also a substantial feature that the coupling between cavity photons
and atoms is generated by an external laser. In the frame rapidly
rotating at the laser frequency,  the time-dependent driving can be
eliminated and the remaining low-frequency dynamics is  described by
the time-independent Dicke-type Hamiltonian
\cite{Nagy2010DickeModel,baumann2010dicke,Ritsch2013Cold}. Note,
however, that the outcoupled field is continuously supplied by the
external laser and a photon current is driven through the system. 

According to the effective Hamiltonian, the incoherent photon population in the ground state diverges at the critical point as a power law with exponent 1/2 \cite{emary2003chaos}. At variance, when the cavity loss is taken into account \cite{nagy2011critical,oztop2012excitations}, the exponent changes to 1 which is much closer to the experimentally measured value of about 0.9 \cite{Brennecke2013Realtime}. The dissipation and the accompanying quantum fluctuations, even at zero temperature, has thus been demonstrated to substantially modify the correlation functions and the critical exponents.

We will show in this Letter that the exponent can be continuously
tuned below 1 and its actual value is determined by the low-frequency
spectral properties of the bath. The value 1 follows from the
Markovian dynamics imposed by photon loss: the `real' frequency of the
leaky photons is in the optical range where the electromagnetic field
has practically a constant spectral density function in the relevant
narrow, say, few kHz bandwidth.  The effect of such a reservoir on the
coupled system can be interpreted effectively as that of a thermal
bath at a non-zero temperature on a single boson mode
\cite{dalla2013keldysh}. However, the bath can have a non-trivial
spectral density function. For example, in the above-quoted
experimental realization there is another dissipation channel acting
on the spin component of the system, which was shown to have a
significant influence on the measured correlation functions
\cite{Brennecke2013Realtime}. Based on a microscopic calculation
beyond the Bogoliubov approach, the observed damping has been
attributed to a Beliaev-type damping process in the superfluid with a
highly non-trivial spectrum of the phonon reservoir at low frequencies
\cite{konya2014photonic,Konya2014Damping}.

We consider a bosonic two-mode model in which one of the modes is
coupled to a simple Markovian bath whereas the other one is subjected
to a colored reservoir. This general model describes with high
accuracy the normal phase of the Dicke-type phase transition
\cite{Nagy2010DickeModel}. The generic feature is that the soft mode
of the phase transition is composed of two bosonic modes which are
coupled by the Hamiltonian
\begin{equation}
\label{eq:dicke}
H_S/\hbar = \omega_a a^\dagger{}a + \omega_b b^\dagger{}b 
+ \frac{y}{2}(a + a^\dagger)(b + b^\dagger)\,.
\end{equation}
The coupling $y$ is restricted to the range below the critical point $y_c$, which we will determine later.  The Keldysh path integral approach is invoked to calculate the dissipative effects \cite{kamenev2011field,dalla2013keldysh}. The action of the Markovian oscillator in frequency space reads
\begin{equation}
\label{eq:Sa}
S_a = \int \frac{\ud\omega}{2\pi}\left(a_{\rm cl}^*, a_{\rm
  q}^*\right) \bmx 0 & \omega-\omega_a-i\kappa
\\ \omega-\omega_a + i\kappa & 2i\kappa \emx \vect{a_{\rm
    cl}}{a_{\rm q}}\,,
\end{equation}
where $a_{\rm cl}(\omega)$ and $a_{\rm q}(\omega)$ are the classical
and quantum fields corresponding to the mode $a$. This mode $a$ is like the `photon' mode in the Dicke model, has a high frequency and $\omega_a$ is referenced to some driving frequency. This mode emits then ``high frequency'' photons
into the vacuum, thus its decay is unaffected by the interaction
between modes $a$ and $b$. Moreover,  the flat reservoir spectrum at high frequencies ensures the validity of a Markovian approximation, which is reflected by using a single constant parameter $\kappa$. It is important to note that the time dependent  external driving, formally eliminated  from the Hamiltonian (\ref{eq:dicke}) in a procedure presented in detail in Ref.~\cite{Ritsch2013Cold}, has as a consequence that a Markovian decay with frequency-independent self-energy function can be used in the above action. 

The genuine low-frequency mode is $b$, which
interacts with a sub-Ohmic bath at zero temperature. It is essential
that this reservoir is sampled at the eigenfrequencies of the coupled  
system. The dissipative action that describes the dynamics reads
\begin{equation}
\label{eq:Seff_b}
S_{b} = \int_\omega 
\left(b_{\rm cl}^*, b_{\rm q}^*\right)
\bmx
 0 & \omega-\omega_b - K^A \\
\omega-\omega_b - K^R & D
\emx
\vect{b_{\rm cl}}{b_{\rm q}}\,,
\end{equation}
where $b_{\rm cl}(\omega)$ and $b_{\rm q}(\omega)$ are the
classical and quantum fields corresponding to mode $b$. The reservoir
is characterized by the coupling-density profile $\rho(\omega)$ which
defines, at zero temperature,  the Keldysh component of the action,  $D(\omega) = 2i\pi{}
\rho(\omega)$, and also the retarded and advanced level-shift functions, 
\be
\label{eq:K_D}
K^{R/A}(\omega) = 
{\cal P}\int_0^\infty\frac{\rho(\omega^{'})}{\omega-\omega^{'}}
 \,\ud\omega^{'} \mp i\pi\rho(\omega) \,.
\ee    
Note the  symmetry \mbox{$K^A(\omega) = [K^R(\omega)]^*$}.  For a sub-Ohmic reservoir, $\rho(\omega)$ starts with an exponent $0 < s < 1$, namely 
\be
\label{eq:rho}
\rho(\omega) =
\Theta(\omega)\frac\gamma\pi\,\frac{\left(\omega/\omega_b\right)^s}{1
  + \left(\omega/\omega_M\right)^2}\,, 
\ee
where $\gamma$ is the dissipation strength, $\omega_M$
is a cutoff frequency, and $\Theta(\omega)$ is the Heavyside function. This reservoir yields a non-conventional level-shift function, which is purely real
for $\omega \leq 0$ and complex for $\omega>0$. The analytical continuation of  $K^R(\omega)$ to the upper half, and $K^A(\omega)$
to the lower half of the complex plane can be fitted together in the function
\be
\label{eq:K_z}
K(z) = \frac{\gamma}{\sin\,\pi{}s}\left(-\frac{z}{\omega_b}\right)^s\,,
\ee
such that it has a branch cut and renders $K^{R/A}(\omega) = \lim_{\eta\rightarrow{}0}K(\omega\pm i\eta)$ on the positive real axis. 

The Keldysh action that corresponds to the interaction term in
Eq.~(\ref{eq:dicke}) is
\begin{equation}
\label{eq:Sab}
S_{ab} = -\frac{y}{2}\int_\omega
\left[(a_q + a_q^*)(b_{cl} + b_{cl}^*) + (a_{cl} + a_{cl}^*)(b_q + b_q^*)\right]\,,
\end{equation}
which contains counter-rotating terms,  like
$a_{cl}b_q$, $a_q^* b_{cl}^*$, etc.   Following the method of
Ref.~\cite{dalla2013keldysh},  the variable space is doubled by introducing fields with negative
frequencies so that the total Keldysh action $S = S_a + S_b + S_{ab}$ can be expressed in a closed quadratic form
\begin{equation}
\label{eq:action_total}
S = \frac12 \int \frac{\ud\omega}{2\pi}
{\bf v}^\dagger
\bmx
 0 & [{\bf G}^A_{4\times{}4}]^{-1}(\omega) \\
[{\bf G}^R_{4\times{}4}]^{-1}(\omega) & {\bf D}^K_{4\times{}4}(\omega)
\emx
{\bf v}\,,
\end{equation}
with the $8$-component field
\begin{multline}
{\bf v}^\dagger(\omega) = [
a_{cl}^*(\omega),
 a_{cl}(-\omega),
b_{cl}^*(\omega),
b_{cl}(-\omega),\\
a_{q}^*(\omega),
a_{q}(-\omega),
b_{q}^*(\omega),
b_{q}(-\omega) ] \,.
\end{multline}
The $4\times{}4$ blocks are matrix Green's functions. The characteristic frequencies of the system correspond to the poles of the retarded Green's function continued analytically to the lower half of the complex plane. The extension is not unique, we use the level-shift function in Eq.~(\ref{eq:K_z}) on the second Riemann sheet
\be
\label{eq:Kz2}
K^{R}_{\rm II}(z) = \gamma\frac{e^{-i \pi{}s}}{\sin\, \pi{}s}(z/\omega_b)^s\,,
\ee
and require the symmetry property  $K^A_{\rm II}(z) = \left[K^R_{\rm  II}(z)\right]^*$.  The poles come then either
in pairs ($z_1$, $-z_1^*$ ),  i.e., opposite real part and the same imaginary part, or  they are purely imaginary.  The 
poles obey the characteristic equation \mbox{${\rm det}[{\bf G}^R_{4\times{}4}]^{-1}(z) = 0$}, i.e., 
\begin{multline}
\label{eq:poles}
[(z+i\kappa)^2-\omega_a^2][(z-i\Gamma(z))^2
  -(\omega_b+\Delta(z))^2] \\ 
 -y^2\omega_a(\omega_b+\Delta(z)) = 0\,, 
\end{multline}
where $\Gamma(z) = (K^R_{\rm II}(z) - K^A_{\rm II}(-z^*))/(2i)$ and $\Delta(z) = (K^R_{\rm II}(z) + K^A_{\rm II}(-z^*))/2$.

\begin{figure}[htb]
\includegraphics[width=0.72\columnwidth]{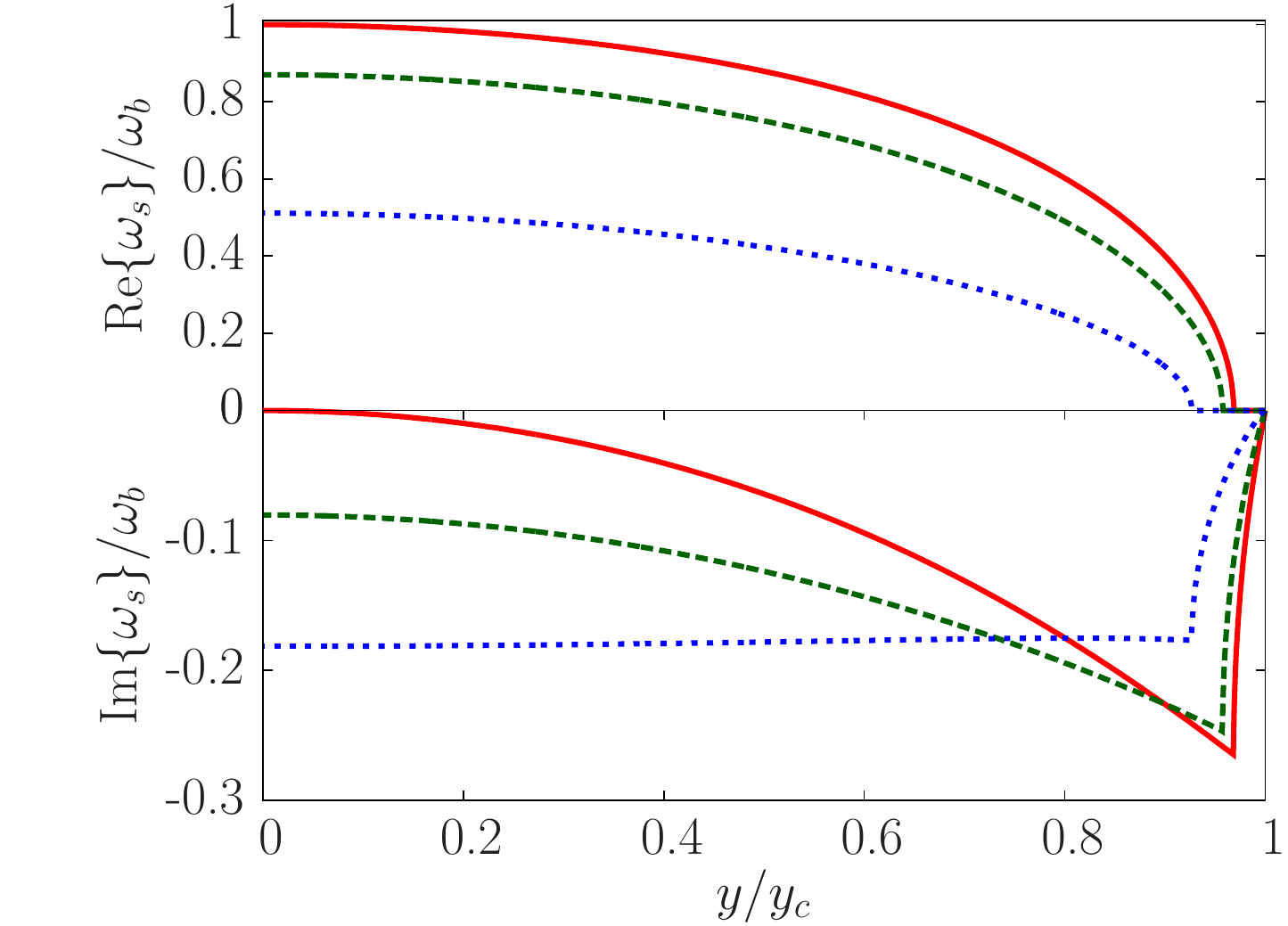}
\caption{(Color online) Real part (upper box) and imaginary part
  (lower box) of the soft mode frequency $\omega_s$ as a function of
  the coupling $y$ to the sub-Ohmic bath for various dissipation
  strengths, $\gamma=0$ (solid red), $0.1$ (dashed green) and
  $0.5\,\omega_b$ (dotted blue). The bath low-frequency exponent is
  $s=4/5$. Other parameters are $\omega_a=\kappa=2\,\omega_b$.}
\label{fig:softmode}
\end{figure}
The numerical solution of this equation for the pole corresponding to
the soft mode frequency $\omega_s$ is shown in Fig.~\ref{fig:softmode}
as a function of the control parameter $y$ for various values of the
reservoir coupling strength $\gamma$. For $\omega_b < \omega_a$, the
soft mode grows, as $y$ is increased from 0, out of the bare mode
$b$. A characteristic feature of second order dissipative phase
transitions is that the real part of the soft mode frequency (top
panel) decreases to zero first, and at this exceptional point a
linewidth bifurcation takes place \cite{Eleuch2013Width}. The larger
the $\gamma$, the real part vanishes for smaller $y$. The critical
point is the one where the upper branch of the imaginary part (bottom
panel) reaches zero,
\begin{equation}
y_c = \sqrt{\frac{\omega_a^2 + \kappa^2}{\omega_a}\omega_b}\,,
\end{equation}
regardless the coupling strength $\gamma$ to the bath.  This is
because the level-shift function vanishes at zero frequency,
$\Gamma(z\rightarrow 0) =0$ and $\Delta(z\rightarrow 0) = 0$, hence it
drops out from the $z=0$ solution of Eq.~(\ref{eq:poles}). The
vanishing of the soft mode frequency implies that there is a
divergence in the correlation functions at $y=y_c$. The exponent,
however, cannot be extracted merely from the position of the poles,
because the dynamics in a coloured reservoir is not
Markovian. Therefore, we resort to a direct, numerically exact calculation of the
correlation functions.

The Fourier-transform of the correlation functions,
\be
 C_{a}(t) = \langle \{a(t),a^\dagger(0)\}\rangle \,, \; C_{b}(t) = \langle \{b(t),b^\dagger(0)\}\rangle\,,
\ee
are given by the components of the Keldysh Green's function, 
$C_{a}(\omega) = i[{\bf G}^K_{4\times{}4}]_{11}$ and $C_{b}(\omega) = i[{\bf G}^K_{4\times{}4}]_{33}$, respectively. 
It can be obtained as
\be 
{\bf  G}^K_{4\times{}4} = -{\bf
  G}^R_{4\times{}4}(\omega){\bf D}^K_{4\times{}4}(\omega){\bf
  G}^A_{4\times{}4}(\omega)\,.
\ee

\begin{figure}[htb]
\includegraphics[width=0.98\columnwidth]{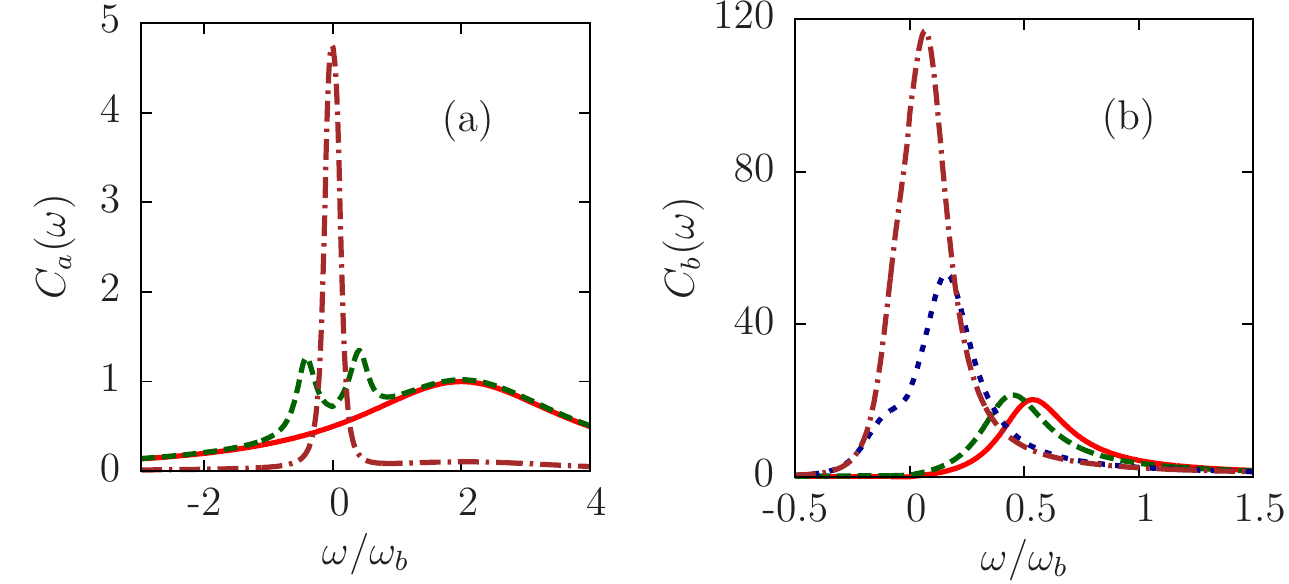}
\caption{(Color online) Power spectrum of the `photonic' mode $a$ (a)
  and of mode $b$ (b) for various coupling strengths approaching the
  critical point: $y=0$ (solid red), $0.5$ (dashed green), $0.9y_c$
  (dotted blue) and $0.95y_c$ (dashed-dotted brown). Parameters:
  $\omega_a=\kappa=2\,\omega_b$, $\gamma=0.5\,\omega_b$, $s=4/5$. 
  In the photonic spectrum (a), the peak for $y=0.95y_c$ is divided by a
  factor of $10$ to fit into the plotted range.}
\label{fig:correl_func}
\end{figure}

Figure~\ref{fig:correl_func} presents the correlation functions
$C_{a}(\omega)$ (a) and $C_{b}(\omega)$ (b) for different values of
the coupling $y$ approaching $y_c$. The former one is measured
directly in the ultracold atom realization of the Dicke model, since
it is the power spectrum of the field leaving the cavity. For a
non-interacting system ($y=0$), the photonic correlation function
(Fig.~\ref{fig:correl_func}a, solid red) is simply a Lorentzian peak
around $\omega = \omega_a$ of width $2\kappa$. It has a non-vanishing part for negative frequencies, causing effectively a heating, which is due to the underlying driving at optical frequency. The correlation
function of mode $b$ for $y=0$ (Fig.~\ref{fig:correl_func}b, solid red) reflects
the properties of the colored bath. This peak is strictly zero for $\omega < 0$ and has a tail only for
large $\omega$, which is consistent with the choice of the coupling
density profile, Eq.(\ref{eq:rho}). Though the spectrum is not Lorentzian, the resonance peak corresponds well to the real part of
the soft-mode frequency, which 
is significantly shifted from the bare frequency $\omega_b = 1$ to $\omega_b = 0.5$ in accordance with Fig.~\ref{fig:softmode} (dotted blue). 

On increasing the coupling $y$, the resonance peak of mode $b$ moves
towards $\omega = 0$, following  the behaviour of the soft-mode in Fig.~\ref{fig:softmode}. The dotted blue line presents the correlation function slightly above the
point where the real part of the soft mode vanishes ($y\approx
0.93y_c$). Such close to the critical point, the correlation function cannot be simply understood in terms of the soft mode.  
This point $y\approx 0.93y_c$ demarcates two different kinds of  correlation function also for the mode $a$, shown in panel (a). Below, there is a doublet symmetrically to $\omega=0$ on top of the broad Lorentzian corresponding to the soft mode frequency shown in
Fig.~\ref{fig:softmode} (the opposite sign eigenfrequency is not shown there). This doublet uncovers the hybridisation
of mode $a$ with mode $b$. For a larger coupling,
e.g. $y=0.95 y_c$ (dashed-dotted brown), a single sharp peak develops in the
origin, that diverges for $y\rightarrow y_c$.

\begin{figure}[t]
\includegraphics[width=0.7\columnwidth]{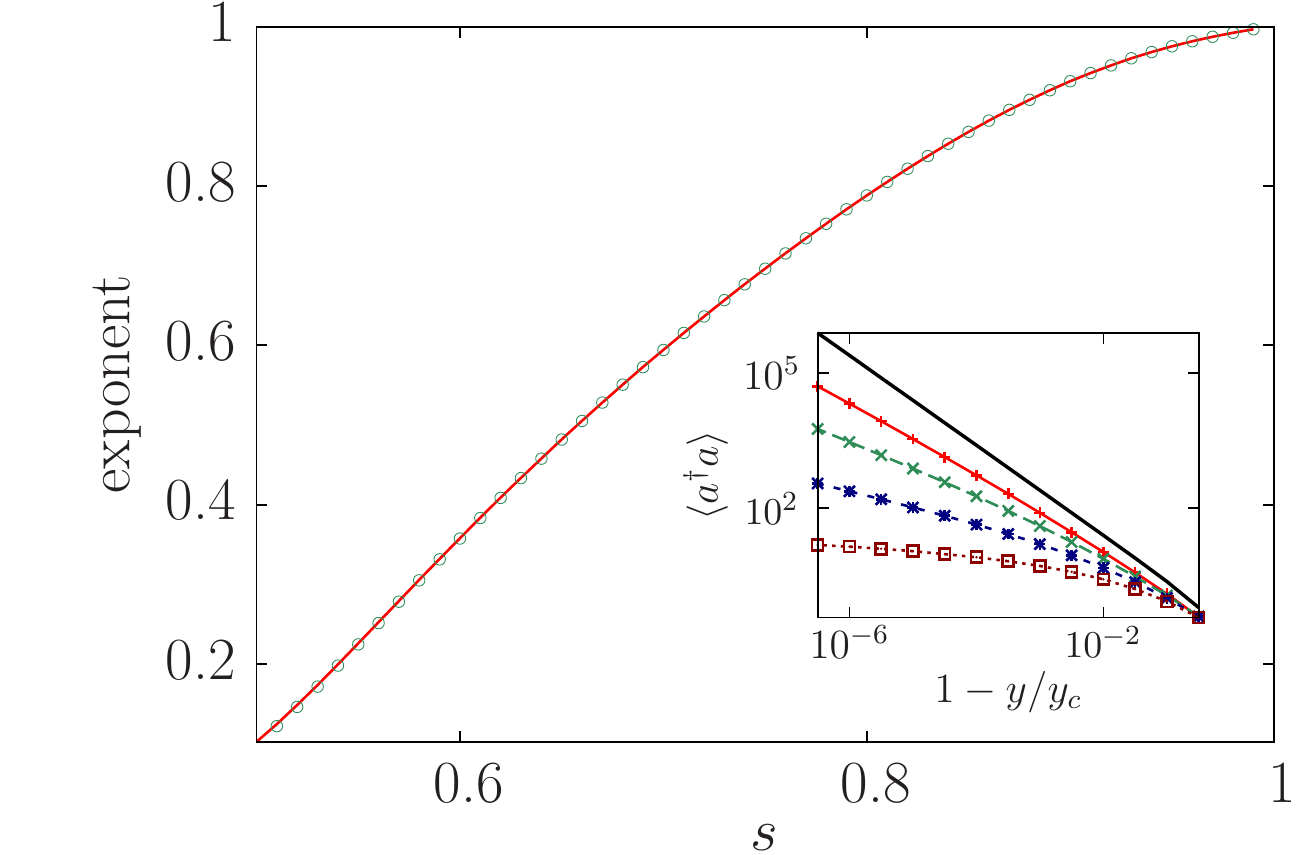}
\caption{(Color online) Critical exponent as a function of the
  exponent $s$ of the sub-Ohmic bath, which is extracted from the
  scaling of the population in mode $a$ (aka `photon number', solid
  red line) and in mode $b$ (green circles). The power-law scaling of
  the photon number is shown in the inset for $s=0.8$ (red crosses +),
  $0.7$ (green crosses x), $0.6$ (blue stars), $0.5$ (brown
  squares). Other parameters are the same as in
  Fig.~\ref{fig:correl_func}. The solid black line shows the $\gamma =
  0$ case for reference.}
\label{fig:phnum_exponent}
\end{figure}

The mean steady-state population $\langle a^\dagger(0) a(0)
\rangle = (C(t=0) - 1)/2$, i.e., cavity photon number, is given by the
integral
\begin{equation}
 \label{eq:correl_func}
C_a(t=0) = \int \frac{\ud\omega}{2\pi} C_a(\omega)\,.
\end{equation}
The photon number diverges at the critical point with a scaling exponent, which is determined numerically. The inset of Fig.~\ref{fig:phnum_exponent} shows the integrated correlation function Eq.~(\ref{eq:correl_func}) for variable coupling $1 - y/y_c$ on a double  logarithmic scale. In the close vicinity of the critical point, $|y - y_c| < 10^{-4}$, the power-law can be recognized and the exponent can be extracted.  The main result of this Letter is summarized in the main panel which presents the continuous and monotonous variation of the  critical exponent as a function of $s$.  For a sub-Ohmic bath the exponent decreases below 1.  Note that the ohmic bath with $s=1$ renders the critical exponent 1 which was known also for the Markovian dissipation without the colored bath ($\gamma=0$) \cite{nagy2011critical,dalla2013keldysh}.

The crucial role of the sub-Ohmic bath in the criticality of
dissipative systems appears also, for example, in the spin-boson model
\cite{Leggett1987Dynamics}, where the exponent of the
localization-delocalization transition depends on the low-frequency
exponent of the spectral function
\cite{Bulla2003Numerical,Winter2009Quantum,Guo2012Critical}.  The
transient (out-of-equilibrium) dynamics exhibits a crossover from the
delocalised to the localised fix points of the equilibrium phase
transition \cite{Anders2007Equilibrium,Kast2013Persistence}.  The
Ising model criticality is also non-trivially influenced by coupling
the spins to a bosonic bath \cite{Werner2005Phase}. In these cases,
the ground state of the coupled system is in question. By contrast,
here we considered an externally driven system in which the
dissipation, more precisely, the interaction with a bath, leads to a
steady state. There is a flow of energy through the coupled $a$ and
$b$ bosons subsystem from the coherent driving into the reservoir,
which cannot be accounted for by an effective Hamiltonian. The
critical point in the steady-state corresponds then to a genuinely
non-equilibrium quantum phase transition
\cite{Schwager2013Dissipative,Lee2013Unconventional}. This Letter was
the first to study non-Markovian bath effects in a non-equilibrium
quantum criticality. We calculated in a numerically exact way the
critical exponent in the open-system Dicke phase transition, and
showed that it is a monotonous function of the low-frequency exponent
of the bath spectrum.

This work was supported by the Hungarian Academy of Sciences (Lend\"ulet Program, LP2011-016).


\begin{thebibliography}{35}
\expandafter\ifx\csname natexlab\endcsname\relax\def\natexlab#1{#1}\fi
\expandafter\ifx\csname bibnamefont\endcsname\relax
  \def\bibnamefont#1{#1}\fi
\expandafter\ifx\csname bibfnamefont\endcsname\relax
  \def\bibfnamefont#1{#1}\fi
\expandafter\ifx\csname citenamefont\endcsname\relax
  \def\citenamefont#1{#1}\fi
\expandafter\ifx\csname url\endcsname\relax
  \def\url#1{\texttt{#1}}\fi
\expandafter\ifx\csname urlprefix\endcsname\relax\def\urlprefix{URL }\fi
\providecommand{\bibinfo}[2]{#2}
\providecommand{\eprint}[2][]{\url{#2}}

\bibitem[{\citenamefont{Morrison and Parkins}(2008)}]{Morrison2008Dynamical}
\bibinfo{author}{\bibfnamefont{S.}~\bibnamefont{Morrison}} \bibnamefont{and}
  \bibinfo{author}{\bibfnamefont{A.~S.} \bibnamefont{Parkins}},
  \bibinfo{journal}{Phys. Rev. Lett.} \textbf{\bibinfo{volume}{100}},
  \bibinfo{pages}{040403} (\bibinfo{year}{2008}).

\bibitem[{\citenamefont{Nagy et~al.}(2008)\citenamefont{Nagy, Szirmai, and
  Domokos}}]{Nagy2008Selforganization}
\bibinfo{author}{\bibfnamefont{D.}~\bibnamefont{Nagy}},
  \bibinfo{author}{\bibfnamefont{G.}~\bibnamefont{Szirmai}}, \bibnamefont{and}
  \bibinfo{author}{\bibfnamefont{P.}~\bibnamefont{Domokos}},
  \bibinfo{journal}{Eur. Phys. J. D} \textbf{\bibinfo{volume}{48}},
  \bibinfo{pages}{127} (\bibinfo{year}{2008}).

\bibitem[{\citenamefont{Piazza and
  Strack}(2014{\natexlab{a}})}]{Piazza2014Quantum}
\bibinfo{author}{\bibfnamefont{F.}~\bibnamefont{Piazza}} \bibnamefont{and}
  \bibinfo{author}{\bibfnamefont{P.}~\bibnamefont{Strack}},
  \bibinfo{journal}{Phys. Rev. A} \textbf{\bibinfo{volume}{90}}
  (\bibinfo{year}{2014}{\natexlab{a}}).

\bibitem[{\citenamefont{Sachdev}(2011)}]{sachdev2011quantum}
\bibinfo{author}{\bibfnamefont{S.}~\bibnamefont{Sachdev}},
  \emph{\bibinfo{title}{Quantum Phase Transitions}}
  (\bibinfo{publisher}{Cambridge University Press}, \bibinfo{year}{2011}), ISBN
  \bibinfo{isbn}{9781139500210}.

\bibitem[{\citenamefont{Strack and Sachdev}(2011)}]{Strack2011Dicke}
\bibinfo{author}{\bibfnamefont{P.}~\bibnamefont{Strack}} \bibnamefont{and}
  \bibinfo{author}{\bibfnamefont{S.}~\bibnamefont{Sachdev}},
  \bibinfo{journal}{Phys. Rev. Lett.} \textbf{\bibinfo{volume}{107}},
  \bibinfo{pages}{277202} (\bibinfo{year}{2011}).

\bibitem[{\citenamefont{Piazza and
  Strack}(2014{\natexlab{b}})}]{Piazza2014Umklapp}
\bibinfo{author}{\bibfnamefont{F.}~\bibnamefont{Piazza}} \bibnamefont{and}
  \bibinfo{author}{\bibfnamefont{P.}~\bibnamefont{Strack}},
  \bibinfo{journal}{Phys. Rev. Lett.} \textbf{\bibinfo{volume}{112}}
  (\bibinfo{year}{2014}{\natexlab{b}}).

\bibitem[{\citenamefont{Dalla~Torre et~al.}(2010)\citenamefont{Dalla~Torre,
  Demler, Giamarchi, and Altman}}]{DallaTorre2010Quantum}
\bibinfo{author}{\bibfnamefont{E.~G.} \bibnamefont{Dalla~Torre}},
  \bibinfo{author}{\bibfnamefont{E.}~\bibnamefont{Demler}},
  \bibinfo{author}{\bibfnamefont{T.}~\bibnamefont{Giamarchi}},
  \bibnamefont{and} \bibinfo{author}{\bibfnamefont{E.}~\bibnamefont{Altman}},
  \bibinfo{journal}{Nature Physics} \textbf{\bibinfo{volume}{6}},
  \bibinfo{pages}{806} (\bibinfo{year}{2010}).

\bibitem[{\citenamefont{Dimer et~al.}(2007)\citenamefont{Dimer, Estienne,
  Parkins, and Carmichael}}]{Dimer2007Proposed}
\bibinfo{author}{\bibfnamefont{F.}~\bibnamefont{Dimer}},
  \bibinfo{author}{\bibfnamefont{B.}~\bibnamefont{Estienne}},
  \bibinfo{author}{\bibfnamefont{A.~S.} \bibnamefont{Parkins}},
  \bibnamefont{and} \bibinfo{author}{\bibfnamefont{H.~J.}
  \bibnamefont{Carmichael}}, \bibinfo{journal}{Phys. Rev. A}
  \textbf{\bibinfo{volume}{75}}, \bibinfo{pages}{013804}
  (\bibinfo{year}{2007}).

\bibitem[{\citenamefont{Nagy et~al.}(2010)\citenamefont{Nagy, K\'{o}nya,
  Szirmai, and Domokos}}]{Nagy2010DickeModel}
\bibinfo{author}{\bibfnamefont{D.}~\bibnamefont{Nagy}},
  \bibinfo{author}{\bibfnamefont{G.}~\bibnamefont{K\'{o}nya}},
  \bibinfo{author}{\bibfnamefont{G.}~\bibnamefont{Szirmai}}, \bibnamefont{and}
  \bibinfo{author}{\bibfnamefont{P.}~\bibnamefont{Domokos}},
  \bibinfo{journal}{Phys. Rev. Lett.} \textbf{\bibinfo{volume}{104}},
  \bibinfo{pages}{130401} (\bibinfo{year}{2010}).

\bibitem[{\citenamefont{Baumann et~al.}(2010)\citenamefont{Baumann, Guerlin,
  Brennecke, and Esslinger}}]{baumann2010dicke}
\bibinfo{author}{\bibfnamefont{K.}~\bibnamefont{Baumann}},
  \bibinfo{author}{\bibfnamefont{C.}~\bibnamefont{Guerlin}},
  \bibinfo{author}{\bibfnamefont{F.}~\bibnamefont{Brennecke}},
  \bibnamefont{and}
  \bibinfo{author}{\bibfnamefont{T.}~\bibnamefont{Esslinger}},
  \bibinfo{journal}{Nature} \textbf{\bibinfo{volume}{464}},
  \bibinfo{pages}{1301} (\bibinfo{year}{2010}).

\bibitem[{\citenamefont{Baumann et~al.}(2011)\citenamefont{Baumann, Mottl,
  Brennecke, and Esslinger}}]{Baumann2011Exploring}
\bibinfo{author}{\bibfnamefont{K.}~\bibnamefont{Baumann}},
  \bibinfo{author}{\bibfnamefont{R.}~\bibnamefont{Mottl}},
  \bibinfo{author}{\bibfnamefont{F.}~\bibnamefont{Brennecke}},
  \bibnamefont{and}
  \bibinfo{author}{\bibfnamefont{T.}~\bibnamefont{Esslinger}},
  \bibinfo{journal}{Phys. Rev. Lett.} \textbf{\bibinfo{volume}{107}},
  \bibinfo{pages}{140402} (\bibinfo{year}{2011}).

\bibitem[{\citenamefont{Mottl et~al.}(2012)\citenamefont{Mottl, Brennecke,
  Baumann, Landig, Donner, and Esslinger}}]{mottl2012roton}
\bibinfo{author}{\bibfnamefont{R.}~\bibnamefont{Mottl}},
  \bibinfo{author}{\bibfnamefont{F.}~\bibnamefont{Brennecke}},
  \bibinfo{author}{\bibfnamefont{K.}~\bibnamefont{Baumann}},
  \bibinfo{author}{\bibfnamefont{R.}~\bibnamefont{Landig}},
  \bibinfo{author}{\bibfnamefont{T.}~\bibnamefont{Donner}}, \bibnamefont{and}
  \bibinfo{author}{\bibfnamefont{T.}~\bibnamefont{Esslinger}},
  \bibinfo{journal}{Science} \textbf{\bibinfo{volume}{336}},
  \bibinfo{pages}{1570} (\bibinfo{year}{2012}).

\bibitem[{\citenamefont{Baden et~al.}(2014)\citenamefont{Baden, Arnold,
  Grimsmo, Parkins, and Barrett}}]{Baden2013Realization}
\bibinfo{author}{\bibfnamefont{M.~P.} \bibnamefont{Baden}},
  \bibinfo{author}{\bibfnamefont{K.~J.} \bibnamefont{Arnold}},
  \bibinfo{author}{\bibfnamefont{A.~L.} \bibnamefont{Grimsmo}},
  \bibinfo{author}{\bibfnamefont{S.}~\bibnamefont{Parkins}}, \bibnamefont{and}
  \bibinfo{author}{\bibfnamefont{M.~D.} \bibnamefont{Barrett}},
  \bibinfo{journal}{Phys. Rev. Lett.} \textbf{\bibinfo{volume}{113}},
  \bibinfo{pages}{020408} (\bibinfo{year}{2014}).

\bibitem[{\citenamefont{Ke\ss{}ler et~al.}(2014)\citenamefont{Ke\ss{}ler,
  Klinder, Wolke, and Hemmerich}}]{Kessler2014Steering}
\bibinfo{author}{\bibfnamefont{H.}~\bibnamefont{Ke\ss{}ler}},
  \bibinfo{author}{\bibfnamefont{J.}~\bibnamefont{Klinder}},
  \bibinfo{author}{\bibfnamefont{M.}~\bibnamefont{Wolke}}, \bibnamefont{and}
  \bibinfo{author}{\bibfnamefont{A.}~\bibnamefont{Hemmerich}},
  \bibinfo{journal}{Phys. Rev. Lett.} \textbf{\bibinfo{volume}{113}},
  \bibinfo{pages}{070404} (\bibinfo{year}{2014}).

\bibitem[{\citenamefont{Schmidt et~al.}(2014)\citenamefont{Schmidt, Tomczyk,
  Slama, and Zimmermann}}]{Schmidt2014Dynamical}
\bibinfo{author}{\bibfnamefont{D.}~\bibnamefont{Schmidt}},
  \bibinfo{author}{\bibfnamefont{H.}~\bibnamefont{Tomczyk}},
  \bibinfo{author}{\bibfnamefont{S.}~\bibnamefont{Slama}}, \bibnamefont{and}
  \bibinfo{author}{\bibfnamefont{C.}~\bibnamefont{Zimmermann}},
  \bibinfo{journal}{Phys. Rev. Lett.} \textbf{\bibinfo{volume}{112}},
  \bibinfo{pages}{115302} (\bibinfo{year}{2014}).

\bibitem[{\citenamefont{Klinder et~al.}(2015)\citenamefont{Klinder, Ke{\ss}ler,
  Wolke, Mathey, and Hemmerich}}]{Klinder2015Dynamical}
\bibinfo{author}{\bibfnamefont{J.}~\bibnamefont{Klinder}},
  \bibinfo{author}{\bibfnamefont{H.}~\bibnamefont{Ke{\ss}ler}},
  \bibinfo{author}{\bibfnamefont{M.}~\bibnamefont{Wolke}},
  \bibinfo{author}{\bibfnamefont{L.}~\bibnamefont{Mathey}}, \bibnamefont{and}
  \bibinfo{author}{\bibfnamefont{A.}~\bibnamefont{Hemmerich}},
  \bibinfo{journal}{Proceedings of the National Academy of Sciences} p.
  \bibinfo{pages}{201417132} (\bibinfo{year}{2015}).

\bibitem[{\citenamefont{Ritsch et~al.}(2013)\citenamefont{Ritsch, Domokos,
  Brennecke, and Esslinger}}]{Ritsch2013Cold}
\bibinfo{author}{\bibfnamefont{H.}~\bibnamefont{Ritsch}},
  \bibinfo{author}{\bibfnamefont{P.}~\bibnamefont{Domokos}},
  \bibinfo{author}{\bibfnamefont{F.}~\bibnamefont{Brennecke}},
  \bibnamefont{and}
  \bibinfo{author}{\bibfnamefont{T.}~\bibnamefont{Esslinger}},
  \bibinfo{journal}{Rev. Mod. Phys.} \textbf{\bibinfo{volume}{85}},
  \bibinfo{pages}{553} (\bibinfo{year}{2013}).

\bibitem[{\citenamefont{Emary and Brandes}(2003)}]{emary2003chaos}
\bibinfo{author}{\bibfnamefont{C.}~\bibnamefont{Emary}} \bibnamefont{and}
  \bibinfo{author}{\bibfnamefont{T.}~\bibnamefont{Brandes}},
  \bibinfo{journal}{Phys. Rev. E} \textbf{\bibinfo{volume}{67}},
  \bibinfo{pages}{066203} (\bibinfo{year}{2003}).

\bibitem[{\citenamefont{Nagy et~al.}(2011)\citenamefont{Nagy, Szirmai, and
  Domokos}}]{nagy2011critical}
\bibinfo{author}{\bibfnamefont{D.}~\bibnamefont{Nagy}},
  \bibinfo{author}{\bibfnamefont{G.}~\bibnamefont{Szirmai}}, \bibnamefont{and}
  \bibinfo{author}{\bibfnamefont{P.}~\bibnamefont{Domokos}},
  \bibinfo{journal}{Phys. Rev. A} \textbf{\bibinfo{volume}{84}},
  \bibinfo{pages}{043637} (\bibinfo{year}{2011}).

\bibitem[{\citenamefont{{\"O}ztop et~al.}(2012)\citenamefont{{\"O}ztop,
  Bordyuh, M{\"u}stecapl{\i}o{\u{g}}lu, and T{\"u}reci}}]{oztop2012excitations}
\bibinfo{author}{\bibfnamefont{B.}~\bibnamefont{{\"O}ztop}},
  \bibinfo{author}{\bibfnamefont{M.}~\bibnamefont{Bordyuh}},
  \bibinfo{author}{\bibfnamefont{{\"O}.~E.}
  \bibnamefont{M{\"u}stecapl{\i}o{\u{g}}lu}}, \bibnamefont{and}
  \bibinfo{author}{\bibfnamefont{H.~E.} \bibnamefont{T{\"u}reci}},
  \bibinfo{journal}{New Journal of Physics} \textbf{\bibinfo{volume}{14}},
  \bibinfo{pages}{085011} (\bibinfo{year}{2012}).

\bibitem[{\citenamefont{Brennecke et~al.}(2013)\citenamefont{Brennecke, Mottl,
  Baumann, Landig, Donner, and Esslinger}}]{Brennecke2013Realtime}
\bibinfo{author}{\bibfnamefont{F.}~\bibnamefont{Brennecke}},
  \bibinfo{author}{\bibfnamefont{R.}~\bibnamefont{Mottl}},
  \bibinfo{author}{\bibfnamefont{K.}~\bibnamefont{Baumann}},
  \bibinfo{author}{\bibfnamefont{R.}~\bibnamefont{Landig}},
  \bibinfo{author}{\bibfnamefont{T.}~\bibnamefont{Donner}}, \bibnamefont{and}
  \bibinfo{author}{\bibfnamefont{T.}~\bibnamefont{Esslinger}},
  \bibinfo{journal}{Proceedings of the National Academy of Sciences}
  \textbf{\bibinfo{volume}{110}}, \bibinfo{pages}{11763}
  (\bibinfo{year}{2013}).

\bibitem[{\citenamefont{Dalla~Torre et~al.}(2013)\citenamefont{Dalla~Torre,
  Diehl, Lukin, Sachdev, and Strack}}]{dalla2013keldysh}
\bibinfo{author}{\bibfnamefont{E.~G.} \bibnamefont{Dalla~Torre}},
  \bibinfo{author}{\bibfnamefont{S.}~\bibnamefont{Diehl}},
  \bibinfo{author}{\bibfnamefont{M.~D.} \bibnamefont{Lukin}},
  \bibinfo{author}{\bibfnamefont{S.}~\bibnamefont{Sachdev}}, \bibnamefont{and}
  \bibinfo{author}{\bibfnamefont{P.}~\bibnamefont{Strack}},
  \bibinfo{journal}{Phys. Rev. A} \textbf{\bibinfo{volume}{87}},
  \bibinfo{pages}{023831} (\bibinfo{year}{2013}).

\bibitem[{\citenamefont{K{\'o}nya et~al.}(2014)\citenamefont{K{\'o}nya,
  Szirmai, Nagy, and Domokos}}]{konya2014photonic}
\bibinfo{author}{\bibfnamefont{G.}~\bibnamefont{K{\'o}nya}},
  \bibinfo{author}{\bibfnamefont{G.}~\bibnamefont{Szirmai}},
  \bibinfo{author}{\bibfnamefont{D.}~\bibnamefont{Nagy}}, \bibnamefont{and}
  \bibinfo{author}{\bibfnamefont{P.}~\bibnamefont{Domokos}},
  \bibinfo{journal}{Phys. Rev. A} \textbf{\bibinfo{volume}{89}},
  \bibinfo{pages}{051601} (\bibinfo{year}{2014}).

\bibitem[{\citenamefont{K\'onya et~al.}(2014)\citenamefont{K\'onya, Szirmai,
  and Domokos}}]{Konya2014Damping}
\bibinfo{author}{\bibfnamefont{G.}~\bibnamefont{K\'onya}},
  \bibinfo{author}{\bibfnamefont{G.}~\bibnamefont{Szirmai}}, \bibnamefont{and}
  \bibinfo{author}{\bibfnamefont{P.}~\bibnamefont{Domokos}},
  \bibinfo{journal}{Phys. Rev. A} \textbf{\bibinfo{volume}{90}},
  \bibinfo{pages}{013623} (\bibinfo{year}{2014}).

\bibitem[{\citenamefont{Kamenev}(2011)}]{kamenev2011field}
\bibinfo{author}{\bibfnamefont{A.}~\bibnamefont{Kamenev}},
  \emph{\bibinfo{title}{Field Theory of Non-equilibrium Systems}}
  (\bibinfo{publisher}{Cambridge University Press}, \bibinfo{year}{2011}).

\bibitem[{\citenamefont{Eleuch and Rotter}(2013)}]{Eleuch2013Width}
\bibinfo{author}{\bibfnamefont{H.}~\bibnamefont{Eleuch}} \bibnamefont{and}
  \bibinfo{author}{\bibfnamefont{I.}~\bibnamefont{Rotter}},
  \bibinfo{journal}{Phys. Rev. E} \textbf{\bibinfo{volume}{87}}
  (\bibinfo{year}{2013}).

\bibitem[{\citenamefont{Leggett et~al.}(1987)\citenamefont{Leggett,
  Chakravarty, Dorsey, Fisher, Garg, and Zwerger}}]{Leggett1987Dynamics}
\bibinfo{author}{\bibfnamefont{A.~J.} \bibnamefont{Leggett}},
  \bibinfo{author}{\bibfnamefont{S.}~\bibnamefont{Chakravarty}},
  \bibinfo{author}{\bibfnamefont{A.~T.} \bibnamefont{Dorsey}},
  \bibinfo{author}{\bibfnamefont{M.~P.~A.} \bibnamefont{Fisher}},
  \bibinfo{author}{\bibfnamefont{A.}~\bibnamefont{Garg}}, \bibnamefont{and}
  \bibinfo{author}{\bibfnamefont{W.}~\bibnamefont{Zwerger}},
  \bibinfo{journal}{Rev. Mod. Phys.} \textbf{\bibinfo{volume}{59}},
  \bibinfo{pages}{1} (\bibinfo{year}{1987}).

\bibitem[{\citenamefont{Bulla et~al.}(2003)\citenamefont{Bulla, Tong, and
  Vojta}}]{Bulla2003Numerical}
\bibinfo{author}{\bibfnamefont{R.}~\bibnamefont{Bulla}},
  \bibinfo{author}{\bibfnamefont{N.-H.} \bibnamefont{Tong}}, \bibnamefont{and}
  \bibinfo{author}{\bibfnamefont{M.}~\bibnamefont{Vojta}},
  \bibinfo{journal}{Phys. Rev. Lett.} \textbf{\bibinfo{volume}{91}}
  (\bibinfo{year}{2003}).

\bibitem[{\citenamefont{Winter et~al.}(2009)\citenamefont{Winter, Rieger,
  Vojta, and Bulla}}]{Winter2009Quantum}
\bibinfo{author}{\bibfnamefont{A.}~\bibnamefont{Winter}},
  \bibinfo{author}{\bibfnamefont{H.}~\bibnamefont{Rieger}},
  \bibinfo{author}{\bibfnamefont{M.}~\bibnamefont{Vojta}}, \bibnamefont{and}
  \bibinfo{author}{\bibfnamefont{R.}~\bibnamefont{Bulla}},
  \bibinfo{journal}{Phys. Rev. Lett.} \textbf{\bibinfo{volume}{102}},
  \bibinfo{pages}{030601} (\bibinfo{year}{2009}).

\bibitem[{\citenamefont{Guo et~al.}(2012)\citenamefont{Guo, Weichselbaum, von
  Delft, and Vojta}}]{Guo2012Critical}
\bibinfo{author}{\bibfnamefont{C.}~\bibnamefont{Guo}},
  \bibinfo{author}{\bibfnamefont{A.}~\bibnamefont{Weichselbaum}},
  \bibinfo{author}{\bibfnamefont{J.}~\bibnamefont{von Delft}},
  \bibnamefont{and} \bibinfo{author}{\bibfnamefont{M.}~\bibnamefont{Vojta}},
  \bibinfo{journal}{Phys. Rev. Lett.} \textbf{\bibinfo{volume}{108}}
  (\bibinfo{year}{2012}).

\bibitem[{\citenamefont{Anders et~al.}(2007)\citenamefont{Anders, Bulla, and
  Vojta}}]{Anders2007Equilibrium}
\bibinfo{author}{\bibfnamefont{F.}~\bibnamefont{Anders}},
  \bibinfo{author}{\bibfnamefont{R.}~\bibnamefont{Bulla}}, \bibnamefont{and}
  \bibinfo{author}{\bibfnamefont{M.}~\bibnamefont{Vojta}},
  \bibinfo{journal}{Phys. Rev. Lett.} \textbf{\bibinfo{volume}{98}}
  (\bibinfo{year}{2007}).

\bibitem[{\citenamefont{Kast and Ankerhold}(2013)}]{Kast2013Persistence}
\bibinfo{author}{\bibfnamefont{D.}~\bibnamefont{Kast}} \bibnamefont{and}
  \bibinfo{author}{\bibfnamefont{J.}~\bibnamefont{Ankerhold}},
  \bibinfo{journal}{Phys. Rev. Lett.} \textbf{\bibinfo{volume}{110}}
  (\bibinfo{year}{2013}).

\bibitem[{\citenamefont{Werner et~al.}(2005)\citenamefont{Werner, V\"olker,
  Troyer, and Chakravarty}}]{Werner2005Phase}
\bibinfo{author}{\bibfnamefont{P.}~\bibnamefont{Werner}},
  \bibinfo{author}{\bibfnamefont{K.}~\bibnamefont{V\"olker}},
  \bibinfo{author}{\bibfnamefont{M.}~\bibnamefont{Troyer}}, \bibnamefont{and}
  \bibinfo{author}{\bibfnamefont{S.}~\bibnamefont{Chakravarty}},
  \bibinfo{journal}{Phys. Rev. Lett.} \textbf{\bibinfo{volume}{94}},
  \bibinfo{pages}{047201} (\bibinfo{year}{2005}).

\bibitem[{\citenamefont{Schwager et~al.}(2013)\citenamefont{Schwager, Cirac,
  and Giedke}}]{Schwager2013Dissipative}
\bibinfo{author}{\bibfnamefont{H.}~\bibnamefont{Schwager}},
  \bibinfo{author}{\bibfnamefont{J.}~\bibnamefont{Cirac}}, \bibnamefont{and}
  \bibinfo{author}{\bibfnamefont{G.}~\bibnamefont{Giedke}},
  \bibinfo{journal}{Phys. Rev. A} \textbf{\bibinfo{volume}{87}},
  \bibinfo{pages}{022110} (\bibinfo{year}{2013}).

\bibitem[{\citenamefont{Lee et~al.}(2013)\citenamefont{Lee, Gopalakrishnan, and
  Lukin}}]{Lee2013Unconventional}
\bibinfo{author}{\bibfnamefont{T.~E.} \bibnamefont{Lee}},
  \bibinfo{author}{\bibfnamefont{S.}~\bibnamefont{Gopalakrishnan}},
  \bibnamefont{and} \bibinfo{author}{\bibfnamefont{M.~D.} \bibnamefont{Lukin}},
  \bibinfo{journal}{Phys. Rev. Lett.} \textbf{\bibinfo{volume}{110}}
  (\bibinfo{year}{2013}).

\end{thebibliography}
\end{document}